\documentstyle{mn}

\newif\ifAMStwofonts

\ifoldfss
  \ifCUPmtlplainloaded \else
    \NewTextAlphabet{textbfit} {cmbxti10} {}
    \NewTextAlphabet{textbfss} {cmssbx10} {}
    \NewMathAlphabet{mathbfit} {cmbxti10} {} 
    \NewMathAlphabet{mathbfss} {cmssbx10} {} 
  \fi
  \ifAMStwofonts
    \ifCUPmtlplainloaded \else
      \NewSymbolFont{upmath} {eurm10}
      \NewSymbolFont{AMSa} {msam10}
      \NewMathSymbol{\upi}     {0}{upmath}{19}
      \NewMathSymbol{\umu}     {0}{upmath}{16}
      \NewMathSymbol{\upartial}{0}{upmath}{40}
      \NewMathSymbol{\leqslant}{3}{AMSa}{36}
      \NewMathSymbol{\geqslant}{3}{AMSa}{3E}

       \let\le=\leqslant
       
    \fi
  \fi
\fi 

\ifnfssone
  \newmathalphabet{\mathit}
  \addtoversion{normal}{\mathit}{cmr}{m}{it}
  \addtoversion{bold}{\mathit}{cmr}{bx}{it}
  \newmathalphabet{\mathbfit} 
  \addtoversion{normal}{\mathbfit}{cmr}{bx}{it}
  \addtoversion{bold}{\mathbfit}{cmr}{bx}{it}
  \newmathalphabet{\mathbfss} 
  \addtoversion{normal}{\mathbfss}{cmss}{bx}{n}
  \addtoversion{bold}{\mathbfss}{cmss}{bx}{n}
  \ifAMStwofonts
    \ifCUPmtlplainloaded \else
      \UseAMStwoboldmath
      \makeatletter
      \new@mathgroup\upmath@group
      \define@mathgroup\mv@normal\upmath@group{eur}{m}{n}
      \define@mathgroup\mv@bold\upmath@group{eur}{b}{n}
      \edef\UPM{\hexnumber\upmath@group}
      \new@mathgroup\amsa@group
      \define@mathgroup\mv@normal\amsa@group{msa}{m}{n}
      \define@mathgroup\mv@bold\amsa@group{msa}{m}{n}
      \edef\AMSa{\hexnumber\amsa@group}
      \makeatother
      \mathchardef\upi="0\UPM19
      \mathchardef\umu="0\UPM16
      \mathchardef\upartial="0\UPM40
      \mathchardef\leqslant="3\AMSa36
      \mathchardef\geqslant="3\AMSa3E

       \let\le=\leqslant

    \fi
  \fi
\fi 

\ifnfsstwo
  \DeclareMathAlphabet{\mathbfit}{OT1}{cmr}{bx}{it}
  \SetMathAlphabet\mathbfit{bold}{OT1}{cmr}{bx}{it}
  \DeclareMathAlphabet{\mathbfss}{OT1}{cmss}{bx}{n}
  \SetMathAlphabet\mathbfss{bold}{OT1}{cmss}{bx}{n}
  \ifAMStwofonts
    \ifCUPmtlplainloaded \else
      \DeclareSymbolFont{UPM}{U}{eur}{m}{n}
      \SetSymbolFont{UPM}{bold}{U}{eur}{b}{n}
      \DeclareSymbolFont{AMSa}{U}{msa}{m}{n}
      \DeclareMathSymbol{\upi}{0}{UPM}{"19}
      \DeclareMathSymbol{\umu}{0}{UPM}{"16}
      \DeclareMathSymbol{\upartial}{0}{UPM}{"40}
      \DeclareMathSymbol{\leqslant}{3}{AMSa}{"36}
      \DeclareMathSymbol{\geqslant}{3}{AMSa}{"3E}

       \let\le=\leqslant

    \fi
  \fi
\fi 

\ifCUPmtlplainloaded \else
  \ifAMStwofonts \else 
    \def\upi{\pi}
    \def\umu{\mu}
    \def\upartial{\partial}
  \fi
\fi

\title[Two Stellar Mass Functions Combined into One]
{Two Stellar Mass Functions Combined into One
by the Random Sampling Model of the IMF}
\author[Bruce G. Elmegreen]
       {Bruce G. Elmegreen\\
BM Research Division, T.J. Watson Research Center,
P.O. Box 218, Yorktown Heights, NY 10598 }
\date{Accepted, Received ;
      in original form }

\pagerange{\pageref{firstpage}--\pageref{lastpage}}
\pubyear{1999}

\begin{document}

\maketitle

\label{firstpage}

\begin{abstract} The turnover in the stellar initial mass function (IMF)
at low mass suggests the presence of two independent mass functions that
combine in different ways above and below a characteristic mass given by
the thermal Jeans mass in the cloud. In the random sampling model
introduced earlier, the Salpeter IMF at intermediate to high mass
follows primarily from the hierarchical structure of interstellar
clouds, which is sampled by various star formation processes and
converted into stars at the local dynamical rate. This power law part is
independent of the details of star formation inside each clump and
therefore has a universal character. The flat part of the IMF at low
mass is proposed here to result from a second, unrelated, physical
process that determines only the probability distribution function for
final star mass inside a clump of a given mass, and is independent of
both this clump mass and the overall cloud structure. Both processes 
operate for all potentially unstable clumps in a cloud, regardless of
mass, but only the first shows up above the thermal Jeans mass, and only
the second shows up below this mass. Analytical and stochastic models of
the IMF that are based on the uniform application of these two functions
for all masses reproduce the observations well. \end{abstract}


\section{Introduction}

Recent theoretical models suggest that the stellar initial mass function
may result from star formation in hierarchically-structured, or
multifractal clouds that are characterized by a local conversion time of
gas into stars that is everywhere equal to the dynamical time on the
corresponding mass scale (Elmegreen 1997, 1999a, hereafter papers I and
II). This model gives the Salpeter power law slope, which was suggested
to be the best representation of the IMF at intermediate to high mass
(see reviews of observations in Scalo 1986, 1998; papers I, II, and Massey
1998), and it gives a break from that slope to a flattened or possibly
decreasing part at lower mass where interstellar clumps cannot become
self-gravitating during normal cloud evolution. This break point may be
at the thermal Jeans mass, given by the Bonner-Ebert critical mass
\begin{equation}
M_{J}=0.35\left({{T}\over{10K}}\right)^{2}\left({{P}\over
{10^6\;k_B}}\right)^{-1/2}\;\;{\rm M}_\odot \label{eq:mj} \end{equation}
for temperature $T$ and total pressure $P$ (thermal + turbulent +
magnetic) in the star-forming region. 

The thermal Jeans mass is fairly constant from region to region in
normal galaxies (Paper II), or at least as constant as the IMF is
observed to be, considering the limited data on the low-mass flattened
part. Yet the characteristic mass $M_J$ may be varied, if needed, to
account for a high mass bias in starburst regions (Rieke et al. 1980) or
the early Universe (Larson 1998) if $T^2$ increases more than $P^{1/2}$.
Such a change might be expected in high density regions because
they have more intense and concentrated star formation. 

A general lack of understanding of the details of star
formation during the final stages has limited the success of the model 
so far to the power law range, and to all of the implications of
stochastic, rather than parameterized, star formation (Papers I and II).
The low mass part of the IMF had not been observed well anyway, so the
original models made no effort to fit any data there.

There is a growing consensus, however, that the low mass IMF becomes
approximately {\it flat} (slope=0) over a significant range in mass when
plotted as a histogram of the log of the number of stars per unit
logarithmic mass interval versus the log of the mass. In such a plot,
the Salpeter (1955) slope is $-1.35$. This flattening was suspected over
two decades ago (Miller \& Scalo 1979), but the most recent observations
are much clearer (Comeron, Rieke, \& Rieke 1996; Macintosh, et al. 1996;
Festin 1997; Hillenbrand 1997; Luhman \& Rieke 1998; Cool 1998; Reid
1998; Lada, Lada \& Muench 1998; Hillenbrand \& Carpenter 1999). 

As a result of these new observations, we suspect that the physical
origin of the low mass IMF can be understood in statistical terms to the
same extent as the power law part. This paper offers one possible
explanation for the flattening of the IMF at low mass that may
be easily tested with modern observations of star formation
in resolved clumps. This explanation
is made here after first reviewing the origin of the power law
part of the IMF at intermediate to high mass. 

\section{The Random Sampling Model}

Diffuse interstellar clouds and the pre-star formation parts of
self-gravitating clouds are generally structured in a hierarchical
fashion with small clumps inside larger clumps over a wide range of
scales (see reviews in Scalo 1985; Elmegreen \& Efremov 2000; for a
review of cloud fractal structure, see St\"utzki 1999). Stellar
masses occur in the midst of this clump range, neither at the smallest
nor the largest ends, and there is no distinction in the
non-self-gravitating (pre-stellar) clump spectrum indicating where the
stellar mass spectrum will eventually lie. These observations
imply that the basic environment
for star formation is scale-free, so the mass scale for stars has
to come from specific physical processes. In papers I and II, we
proposed that the mass scale arises because of the need for
self-gravity to dominate the most basic of all
forces, thermal pressure, and this leads to the $M>M_J$ constraint
discussed above.  In fact, the break point in the power law part
of the IMF arises at about $M_J$ for normal conditions.

The Salpeter power-law portion of the IMF was then proposed to
arise as stars form out of gas structures that lie at random
levels in the hierarchy (where $M>M_J$). Perhaps the physical process
that initiates this is turbulence compression followed by gravitational
collapse of the compressed slab (Elmegreen 1993; Klessen, Heitsch, \&
MacLow 2000), or perhaps it is an external compression acting on a
certain gas structure. Regardless, the star formation process {\it
samples} the hierarchical structure of the cloud in this model, and
builds up a stellar mass spectrum over time. 

This process of {\it sampling} where and when a particular star forms
can only be viewed as {\it random} at the present time, just as the time
and place of rain in a terrestrial cloud pattern is random and given
only as a probability in most weather forecasts. The detailed star
formation processes are not proposed to be random, only the
manifestation of them, considering that turbulence is too complex for
initial and boundary conditions to be followed very far over time
and space with enough certainty to produce a predictable result. Thus we
discuss here and in the previous two papers only the probability of
sampling from various hierarchical levels, and we use this probability
to generate a stochastic model that builds up the IMF numerically after
many random trials. From this point of view, the basic form of the IMF
is fairly easy to understand, even though the final slope has not been
derived analytically (i.e., it has been obtained only by running the
computer model).

For example, if the sampling process has a uniform probability over all
hierarchical levels, then the mass spectrum is $M^{-2}dM=M^{-1}d\log M$
exactly (e.g., Fleck 1996; Elmegreen \& Falgarone 1996). A dynamically
realistic cloud will not sample in such a uniform way, however. It will
be biased towards regions of higher density as they evolve more quickly.
For most dynamical processes that precede star formation, including
self-gravitational contraction, magnetic diffusion, and turbulence, the
rate of evolution scales with the square root of the average local
density. This means that lower mass clumps are sampled more often than
higher mass clumps. As a result, the mass spectrum steepens to
$M^{-1.15}d\log M$. The extra $-0.15$ in the power comes from $(D-3)/2D$
for fractal dimension $D$ (Paper I).

In addition to this steepening from density weighting, there is another
steepening effect from mass competition. Once a small mass clump turns
into an independent, self-gravitating region, the larger mass clump that
surrounds it has less gas available to make another star. If it ever
does make a star, then the mass of this star will be less than it would
have been if the first star inside of it had never formed. After these
two steepening effects, the IMF becomes $M^{-1.35}d\log M$, which is the
Salpeter function.  

\section{The Flat Part of the IMF}

The process of random, weighted sampling from hierarchical clouds
changes below the thermal Jeans mass because most of the clumps there
cannot form stars at all: there is not enough self-gravity to give
collapse no matter how the clump is put together. As a result, the
process of star formation stops at sufficiently low clump mass. This
does not mean that stars smaller than $M_J$ cannot form. Each collapsing
clump turns into one or more stars with some efficiency less than unity,
so although the average star mass is proportional to the clump mass in
this model, the actual star mass can be considerably less. There should
even be a range of stellar masses coming from each self-gravitating
clump of a given mass, depending on how the clump divides itself into
stars and how much disk and peripheral gas gets thrown back without
making stars. 

We have now discovered a curious aspect to this random sampling model
when the conversion of clump mass into star mass is considered
explicitly. That is, the probability distribution function for this
conversion reveals itself clearly below the characteristic mass, and is,
in fact, identical to the observed form of the IMF there. We also find
that this clump-to-star distribution function can apply equally well
above and below $M_J$, in a self-similar way, but that it {\it only}
shows up below this mass. Thus the power law part of the IMF is nearly
independent of the details of how a clump gets converted into stars (as
long as the dynamical rate is involved) and depends primarily on the
cloud structure, whereas the flat part of the IMF depends exclusively on
the details of clump-to-star conversion and is independent of the nature
of the cloud's structure. 
 
To describe this process mathematically, we suppose that each
self-gravitating clump of mass $M_c$ makes a range of star masses $M_s$
such that the probability distribution function, $P(\epsilon)$, of the
relative star mass, $\epsilon=M_s/M_c$, is independent of $M_c$. This is
consistent with the self-similarity of star formation that is assumed
for the rest of the IMF model. The distribution function is written for
logarithmic intervals as $P(\epsilon)d\log\epsilon$. The basic point of
this paper is that $P(\epsilon)$ must be approximately constant for all clump
masses to give the flat part of the IMF at low mass.

The final mass function for stars, in logarithmic intervals,
$n_s(M_s)d\log M_s$, can now be written in terms of the mass
function for self-gravitating, randomly-chosen clumps, 
$n_c(M_c)d\log M_c$, as
\begin{equation}
n_s\left(M_s\right)=\int_{\epsilon_{min}}^{\epsilon_{max}}
P(\epsilon)n_c\left(M_s/\epsilon\right)d\log\epsilon .
\label{eq:int}\end{equation}
The upper limit to the integral, $\epsilon_{max}$, is the largest
relative mass for a star that is likely to form from a self-gravitating clump.
It is perhaps slightly less than unity when $M_s>>M_J$, although its
precise value is not necessary here.  We denote it by the constant
$\epsilon_{max,0}$
and take $\epsilon_{max}=\epsilon_{max,0}$ when $M_s>M_J\epsilon_{max,0}$.
When $M_s<M_J\epsilon_{max,0}$, the efficiency can be at most
$M_s/M_J$ since the smallest clump that can form stars is $M_J$. Thus
\begin{equation}
\epsilon_{max}={{M_s}\over{MAX\left(M_J,M_s/\epsilon_{max,0}\right)}}.
\label{eq:emax}\end{equation}

The lower limit to the integral in equation (\ref{eq:int}),
$\epsilon_{min}$, is not known yet from observations. The ratio of
$\epsilon_{max,0}$ to $\epsilon_{min}$ will turn out to be the mass
range for the flat part of the IMF, which is denoted by ${\cal R}$.
It may have a value of $\sim10$ or more according to recent
observations (Sect. I), but future observations of lower mass brown
dwarfs could extend the flat part further and make ${\cal R}$ larger.
In any case, we take
\begin{equation}
\epsilon_{min}=\epsilon_{max,0}/{\cal R};\label{eq:emin}\end{equation}
$\epsilon_{min}$ does not depend on whether $M_s$ is greater
than or less than $M_J\epsilon_{max,0}$. 

To solve the integral in equation (\ref{eq:int}), we convert
$\epsilon$ to $M_s/M_c$ and $d\log\epsilon=-d\log M_c=-dM_c/M_c$
for constant $M_s $ from the left hand side. Then, for
logarithmic intervals in $M_s$, 
\begin{equation}
n_s(M_s)=P_0n_0\int_{M_{c,min}}^{M_{c,max}} M_c^{-1-x}dM_c
\label{eq:int2}\end{equation}
where $P_0=P(\epsilon)=1/\int d\log\epsilon=
1/\log{\cal R}$ for all
$M_c$, and $n_0$ comes from the constant in the clump function,
$n(M_c)=n_0M_c^{-x}$; $x=1.35$ from the random sampling
model and is the same as the power law in the Salpeter function.
The integral limits are $M_{c,min}=M_s/\epsilon_{max}=MAX\left(M_J,
M_s/\epsilon_{max,0}\right)$ and $M_{c,max}=M_s/\epsilon_{min}
=M_s{\cal R}/\epsilon_{max,0}$.

The solution to equation (\ref{eq:int2}) depends on whether 
$M_J\epsilon_{max,0}$ is greater than or less than $M_s$.
For $M_s>M_J\epsilon_{max,0}$,
\begin{equation}
n_s(M_s)={{P_0n_0}\over{x}}\left({{\epsilon_{max,0}}
\over{M_s}}\right)^x
\left(1-{\cal R}^{-x}\right)\propto M_s^{-x} .
\label{eq:sol1}
\end{equation}
For $M_J\epsilon_{max,0}/{\cal R}<M_s<M_J\epsilon_{max,0}$,
\begin{eqnarray}
n_s(M_s)={{P_0n_0}\over{xM_J^x}}
\left[1-\left({{M_J\epsilon_{max,0}}\over{M_s{\cal R}}}\right)^{x}\right]
\approx \;\;{\rm constant},
\label{eq:sol2}
\end{eqnarray}
and for 
$M_s\le M_J\epsilon_{max,0}/{\cal R}$, this latter result is zero.
At $M_s=M_J\epsilon_{max,0}$, the expressions in equations
(\ref{eq:sol1}) and (\ref{eq:sol2}) are the same. 
A graph of $n_s(M_s)$ from these equations is shown in figure 1 for
${\cal R}=10$, $\epsilon_{max,0}=1$, and $M_J=0.3$ M$_\odot$.

\begin{figure}
\vspace{3.in}
\includegraphics{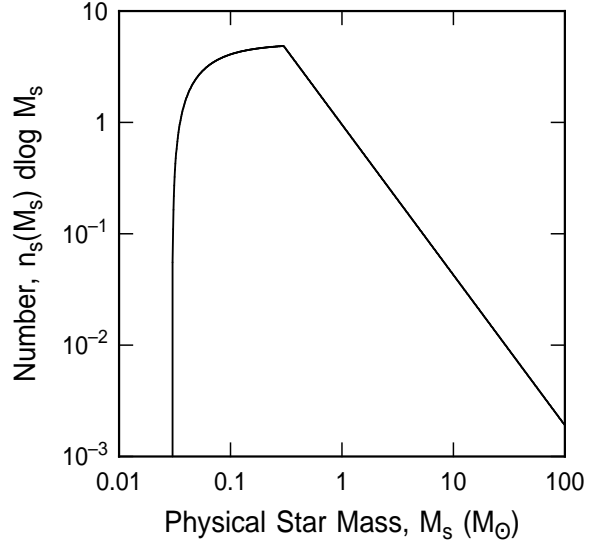}
\caption{Solution to the analytical expression for the IMF
from equations 6 and 7 using ${\cal R}=10$.}
\end{figure}
\begin{figure}
\vspace{5.3in}
\includegraphics{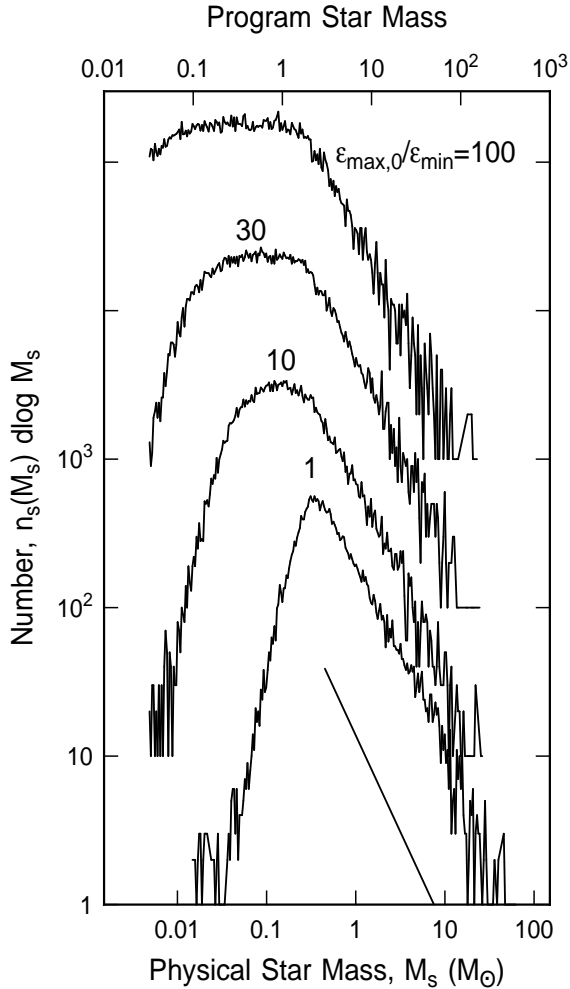}
\caption{IMF model with random sampling from a
hierarchical cloud for four values of the mass range ${\cal R}$,
defined to be the ratio of the maximum fraction to the minimum 
fraction of the clump mass that goes into a star. The straight line
has a slope of 1.3}
\end{figure}

Figure 2 shows the result better using the numerical model with random
sampling described in Papers I and II, but now with a stellar mass equal
to a random fraction $\epsilon$ of the chosen clump mass. This random
fraction is equally distributed over a logarithmic range, as specified
in the above discussion, by using the equation $\epsilon= {\cal R}^{r-1}$ for
random variable $r$ that is distributed uniformly over
$\left[0,1\right]$. Four different values of ${\cal R}$ are used in the
figure to show how the length of the flattened part equals ${\cal R}$.
The ${\cal R}=1$ case shows the pure clump selection spectrum, as
in papers I and II, but with a sharper lower mass cutoff than in the
other papers, taken here
from a failure probability $P_f=exp\left(-\left[M_c/M_J\right]^{4}\right)$.
We also took $\epsilon_{max,0}=1$ for simplicity; the exact value is not
important (it occurs in the expression for $\epsilon(r)$). 
For all of the cases, there are 10 levels of hierarchical
structure in the cloud model, with an average of 2 subpieces per piece,
and an actual number of subpieces per piece distributed as a Poisson
variable over the interval from 1 to 4.

The results indicate that for $M_s>M_J\epsilon_{max,0}$, the model IMF
is a power law with the same power as the clump mass spectrum obtained
previously. The distribution $P\left(\epsilon\right)$ does not affect
the IMF above the break point even though it applies there. This is
because each chosen clump makes a wide range of stellar masses and
contributes a flat component of width ${\cal R}$ to the local IMF, but
the {\it sum} of the number of stars at each stellar mass is still a
power law with the same power as the clump spectrum, independent of
$P\left(\epsilon\right)$. 

The IMF becomes flat below
$M_J\epsilon_{max,0}$ because all of the stars there come from 
clumps with masses near $M_J$. In fact, the decreasing nature of the clump
spectrum toward higher masses makes clumps with masses very close to
$M_J$ the favored parents for stars with masses less than 
$M_J\epsilon_{max,0}$.  Thus the low mass IMF
is determined entirely by the {\it
separate} mass spectrum for stars that form inside each self-gravitating
clump.

Evidently, there are two mass spectra for star formation, one coming
from cloud structure and clump selection, giving the Salpeter function,
and another coming from star formation inside each clump, giving the
flat part. The latter function actually applies everywhere, but it does
not visibly affect the Salpeter distribution for intermediate to high
masses because stars there have a wide range ($={\cal R}$) of clump
masses for parents. It only appears for star masses that form from the
lowest mass clumps that can make a star. 

The model IMF decreases sharply below the smallest star mass that can
form in a clump of mass $M_J$, which is
$M_J\epsilon_{max,0}/{\cal R}$. 
The parameters $\epsilon_{max,0}$ and ${\cal
R}$, as well as the function $P(\epsilon)$, depend on the physics of
star formation, unlike $x$ in the power law part of the IMF, which
depends primarily on the physics of prestellar cloud structure. 

\section{Conclusions}

The flat part of the IMF is proposed to result from the distribution of
the ratio of star mass to clump mass, which must also be flat in
logarithmic intervals. Such a distribution shows up only below the mass
of the smallest unstable clump mass, so in principle, it might only
apply there physically.  However, it could apply equally well for all
clumps in a star-forming cloud because it does not actually reveal
itself at masses above this threshold. The low mass flattening and
eventual turnover below the brown dwarf range depend in unspecified ways
on the detailed physics of star formation, which also contribute to the
single characteristic mass, $M_J$. The slope of the power law part of
the IMF, above $M_J$, depends primarily on cloud structure, although the
same complexities of star formation probably apply there too.
This is why the Salpeter IMF appears in so many diverse environments: the
universal character of turbulent cloud structure determines the power
law nature and the power law itself for intermediate to high mass stars,
independent of the details of star formation.
Variations in the IMF from diverse physical conditions
should be much more pronounced at low mass.

This model may be checked observationally by determining the range and
distribution function for final star mass in resolved clumps of a given
mass. The resolution of the clumps is important so one can be sure the
clumps belonging to each individual star or binary pair are measured,
and not confused with larger clumps that may include substructure
associationed with separate stars. If this model is correct, then the
star mass distribution for clumps of a given mass will be much flatter
than the Salpeter function, and this may be the case for all clump
masses, even those containing intermediate to high mass stars where the
Salpeter function applies to the ensemble of stars coming from clumps of
different masses.

\bsp

\label{lastpage}

\end{document}